# Modeling of scalar dissipation rates in flamelet models for low temperature combustion engine simulations

(Full-Length Article)

*Saurabh Gupta[a] (guptasrb@umich.edu), SeungHwan Keum[b] (sgtkeum@umich.edu), Pinaki Pal[a] (pinaki@umich.edu), Hong G. Im[c,*](hong.im@kaust.edu.sa)*

[a]Department of Mechanical Engineering, University of Michigan, 2350 Hayward Street, Ann Arbor, MI 48109-2125, USA

[b]General Motors R&D PSR Laboratory, 30500 Mount Rd, Warren, MI 48090, USA

[c]Clean Combustion Research Center, King Abdullah University of Science and Technology, Thuwal 23955-6900, Saudi Arabia

*Corresponding Author:

Hong G. Im
Clean Combustion Research Center
King Abdullah University of Science and Technology
Thuwal 23955-6900, Saudi Arabia

**ABSTRACT**

The flamelet approach offers a viable framework for combustion modeling of homogeneous charge compression ignition (HCCI) engines under stratified mixture conditions. Scalar dissipation rate acts as a key parameter in flamelet-based combustion models which connects the physical mixing space to the reactive space. The aim of this paper is to gain fundamental insights into turbulent mixing in low temperature combustion (LTC) engines and investigate the modeling of scalar dissipation rate. Three direct numerical simulation (DNS) test cases of two-dimensional turbulent auto-ignition of a hydrogen-air mixture with different correlations of temperature and mixture fraction are considered, which are representative of different ignition regimes. The existing models of mean and conditional scalar dissipation rates, and probability density functions (PDFs) of mixture fraction and total enthalpy are *a priori* validated against the DNS data. The underlying reasons behind the performance of the existing models are explored by looking into the local and global dynamics of mixing. In the context of Reynolds-averaged Navier-Stokes (RANS) framework, the current models are not able to capture the local rise in scalar dissipation rate due to the effects of differential diffusion of hydrogen. However, the beta PDF gives reasonably accurate results when differential diffusion effects are discarded. It is also observed that model constants for the mean scalar dissipation rates should be taken as 3.0 instead of 2.0 for more accurate results. In the context of large eddy simulation (LES), it is found that mixing is completely uncorrelated with turbulence and that the mean scalar dissipation rates can be accurately modeled as being proportional to the scalar variances. Based on this understanding, novel simplistic models for the mean and cross scalar dissipation rates are proposed and are shown to perform very well under the wide range of auto-ignition regimes considered in this study.

**Short title:** Scalar dissipation rates in flamelet models

# 1. INTRODUCTION

The ongoing demands for internal combustion (IC) engines with higher efficiency and lower emissions have attracted research interests in low temperature combustion (LTC) of dilute premixed or partially premixed fuel-oxidizer mixtures. Homogeneous charge compression ignition (HCCI) is one of the earlier developments utilizing the LTC concept, and has been extensively investigated over the last decades [1,2] showing promises in lower $NO_x$ and soot emissions and higher efficiencies relative to the conventional spark ignition (SI) and compression ignition (CI) engines. For such a nearly homogeneous combustion system, full-cycle engine simulations based on the multi-zone combustion submodels have been conducted with great success [3-6]. Practical implementation of the HCCI engines, however, has faced challenges in the accurate control of ignition timing and combustion phasing, thereby restricting the applicable range of load conditions. To overcome the difficulties, a large number of modified LTC approaches have been proposed, either by utilizing exhaust/residual gas mixing through variable valve timing control [7], by multi-stage fuel injection [8], by the spark assistance [9], or by dual fuel usage [10]. A common characteristic of the different renditions of modern LTC engines is that they all exploit certain degrees of mixture inhomogeneity and the resulting mixed-mode combustion: premixed, non-premixed, or nearly homogeneous (ignition-dominant) modes. As such, predictive modeling of modern IC engines should properly account for the complex combustion characteristics arising from the presence of mixture stratifications.

In the present study, the flamelet model [11] is adopted as the main conceptual framework of LTC combustion sub-model development and refinement. The mathematical basis of the flamelet approach is that the evolution of reactive species and chemical reactions is described in a reduced dimensional space spanned by one or two conserved scalar variables. Consequently, the multi-dimensional computational fluid dynamics (CFD) simulations, which constitute a major portion of the computational cost, need only to solve non-reactive conserved variables. The method has been successfully applied to diesel combustion applications [12, 13] and has been extended to incorporate multiple-injection strategies [14]. Considering the underlying principle, however, the flamelet approach should be applicable to a broader range of mixed-mode combustion regimes. In particular, the representative interactive flamelet (RIF) approach [11], which solves the unsteady flamelet equation in the reactive space in conjunction with the CFD time integration, is considered well suited to describe the important transient characteristics in LTC operations, such as auto-ignition and subsequent front propagation. Our recent study has implemented a spray-interactive flamelet formulation in order to capture the late injection combustion conditions more accurately [15]. A series of fundamental studies have been conducted through direct numerical simulations (DNS) in HCCI-like environment [16-19] and the validity of the RIF approach has been demonstrated [16, 19]. The present study follows a similar methodology in utilizing DNS in *a priori* validation, but with an emphasis on the mixing model.

In the RIF approach, the mixture fraction variable, *Z*, often defined by the Bilger formula [20], is commonly used as the one-dimensional conserved scalar. To account for non-uniform enthalpy levels due to exhaust gas mixing and wall heat loss, the total enthalpy variable, *H*, can be added to construct a two-dimensional flamelet equation in the *Z-H* space [19, 21]. In this formulation, the important parameters coupling the physical (CFD) and reactive (flamelet) spaces are the scalar dissipation rates for both *Z* and *H*, defined as

$$\chi_Z = 2\alpha |\nabla Z|^2, \quad \chi_H = 2\alpha |\nabla H|^2 \qquad (1)$$

where *α* is commonly chosen to be the local thermal diffusivity. At each time step of the numerical integration, the equations in the physical space are first solved to obtain the mean scalar dissipation rates at stoichiometric conditions, $\tilde{\chi}_{Z,st}$ and $\tilde{\chi}_{H,st}$. This is then used to calculate the conditional scalar dissipation rates, $\langle c_Z | Z \rangle$ and $\langle c_H | H \rangle$, which are the key parameters in the flamelet equations. The relations between the two variables are commonly determined in by an analytical function. Once the reactive scalar variables are updated in the flamelet space, their mean values in the CFD cells are determined by the probability density function (PDF) integrals. The specific form of the PDF function is often presumed in terms of the CFD solution variables, such as mean and variance of the conserved scalar variables, $Z$ and $H$. While the framework and the specific functional forms have been well established, they were developed based on the consideration of non-premixed combustion, and thus their validity in the mixed-mode combustion conditions needs to be carefully examined.

Therefore, the objectives of the present study are to assess the fidelity of various existing mixing models employed in the flamelet formulation in their application to LTC ignition/combustion cases. In particular, the primary quantities to be evaluated are:

- The probability density functions of $Z$ and $H$
- The mean $Z$ and $H$ scalar dissipation rates
- The conditional scalar dissipation rates for $Z$ and $H$

The *a priori* study will be conducted by utilizing two-dimensional DNS of auto-ignition in the presence of temperature and composition fluctuations, taken from a previous study [22]. The analysis is done in the context of both RANS and LES formalisms.

Section 2 of this paper summarizes the initial conditions and numerical setup used in the DNS study. In Section 3, the existing model is validated and its limitations are identified, and a new and improved model for the mean $Z$ and $H$ scalar dissipation rates is proposed in the context of RANS formulation. Subsequently, the cross scalar dissipation rates are investigated and an additional model improvement is suggested. Finally, the performance of the existing models in the context of LES is examined. The key findings from this study are summarized in the conclusions section.

## 2. DESCRIPTION OF SIMULATION

To conduct an *a priori* test, two-dimensional DNS of auto-ignition of a turbulent $H_2$-air mixture in a closed volume in the presence of inhomogeneous initial composition and temperature fields are used for validation [22]. The initial conditions include a mean temperature of 1070 K, mean hydrogen/air equivalence ratio of 0.1, a uniform pressure of 41 atm and a turbulent Reynolds number of 51. Further details of the numerical algorithms, initialization and the grid resolution can be found in Ref. [22]. The DNS used a detailed $H_2$/air chemical kinetic mechanism with 9 species and 22 chemical reactions developed by Mueller et al. [23]. The simulations considered the following three cases in terms of mixture inhomogeneities imposed as the initial condition:

- Temperature inhomogeneities only, in which combustion is dominated by spontaneous front propagation (Case A)

- Uncorrelated temperature and compositional inhomogeneities, in which combustion is dominated by deflagrative front propagation (Case B)

- Negatively-correlated temperature and compositional inhomogeneities, in which combustion is dominated by nearly homogeneous auto-ignition (Case C)

Therefore, the DNS data serves as a good reference to test the flamelet models for a wide range of LTC engine operation conditions.

## 3. RESULTS AND DISCUSSION

### 3.1. Assessment of RANS Mixing Models

*3.1.1. Probability density functions for Z and H*

In what follows, the mixture fraction and enthalpy variables are normalized based on their minimum and maximum values at each time step, such as

$$Z = \frac{Z - Z_{min}}{Z_{max} - Z_{min}} \tag{2}$$

where ζ is the actual mixture fraction quantity at a given time, ranging from the minimum and the maximum values. The normalized mixture fraction and enthalpy variables are actually solved in the flamelet equation [24]. At each time of the DNS calculations, the solution field is processed to compute the PDF for the mixture fraction and enthalpy variables, and the results are compared with the presumed beta-PDF.

Figures 1(a) and (b) show the PDF of $Z$ at four different times during the evolution, for Cases B and C, respectively. Cases B and C represent ignition event dominated by front propagation and nearly homogeneous volumetric ignition [22], thus the two cases cover a wide range of LTC operation conditions. PDFs of $H$ follow a similar trend and have not been shown here. Comparison between the two cases shows an important difference. For Case B (Fig. 1(a)), the peak in the PDF curve shifts towards a lower $Z$ value up to 2 ms, and then shifts towards a higher $Z$ value later in time. On the other hand, no such shifts in the peak location are observed for Case C (Fig. 1(b)). This suggests that there is a difference in the mixing processes between the two cases.

To investigate the cause of this behavior, Figure 2 shows the evolution of mixture fraction field for Case B. The ignition fronts form at around $t = 1.5$ ms, creating a sudden increase in the mixture fraction gradient as the ignition front separates the unburned and the burned regions. Although not shown here, the $H$ field also exhibits similar behavior. Therefore, while it may seem contradictory because a conserved scalar variable should by definition remain conserved through combustion, it is evident that ignition and front generation creates strong gradients in the mixture fraction variable, and the corresponding increase in the scalar dissipation rate, which causes a modeling challenge as will be discussed later. This is coupled to the differential diffusion process to yield the shift in the PDF peak. As the ignition fronts form, hydrogen from the unburned mixture diffuses to front at a much faster rate than other species, hence creating a lower local *Z* level in the affected unburned mixture region, thus leading to the shift in the PDF peak as shown in Figure 1(a). A similar observation is found for the total enthalpy variable since hydrogen carries a larger value of chemical enthalpy. After 2 ms, as the ignition front spreads out and the reactant mixture is depleting, stronger gradients in *Z* vanishes and thus the peak scalar dissipation rate shifts back to the original shape.

In contrast, Figure 3 shows the evolution of the mixture fraction field for Case C. Since the ignition front formation is hardly present, there is no prominent occurrence of generation in the mixture fraction gradient, and thus a shift in the PDF peak is not as observed in Figure 1(b).

The PDF curves generated from the actual DNS data are compared with the beta-PDF, determined by [11, 25]:

$$P(Z) = \frac{Z^{a-1}(1-Z)^{b-1}}{\Gamma(a)\Gamma(b)} \Gamma(a+b) \tag{3}$$

$$a = \tilde{Z}\gamma, \; \beta = (1-\tilde{Z})\gamma, \; \gamma = \frac{\tilde{Z}(1-\tilde{Z})}{\widetilde{Z''^2}} - 1 \tag{4}$$

where $\Gamma$ is the gamma function. PDFs of *H* follow a similar trend and have not been shown here. Figures 4(a) and (b) show the beta PDF evolution for case B and case C, respectively, at the same times shown in Figure 1. A good agreement is found between the DNS and the beta PDF model, especially for Case C. For Case B, however, the beta-PDF model underpredicts the DNS results at 2 ms and 2.5 ms. Again, this is attributed to the occurrence of small length scale ignition fronts creating large local gradients in *Z* and *H*. Since beta-PDF formula, Equations (3) and (4), is based on the volumetric mean and variance of *Z* (and *H*), it is unable to capture the small-scale features resulting from the front formation.

To confirm that the shift in the PDF peak is indeed caused by the differential diffusion effect, a contrived simulation is also conducted by setting the same condition as Case B, except that the Lewis number of each species is set to unity so that the differential diffusion effect is eliminated. Figure 5 shows the result for both DNS and the beta-PDF calculations, and it is clearly seen that the peak shift is not observed. Furthermore, the overall magnitude of the PDF curves between the DNS and beta-PDF results agree more closely. Overall, the beta-PDF model is found to be reasonable in describing LTC ignition and combustion processes, provided the differential diffusion effect can be properly accounted for.

*3.1.2. Reynolds-averaged scalar dissipation rates*

The Reynolds-averaged scalar dissipation rates, $\tilde{\chi}_Z$ and $\tilde{\chi}_H$, appear as unclosed terms in the transport equations for the variance, $Z''^2$ and $H''^2$ [11]. A commonly adopted approach is the constant mixing-to-turbulent timescale model [11, 24]:

$$\tilde{\chi}_Z = C_Z \frac{\tilde{\varepsilon}}{\tilde{\kappa}} \widetilde{Z''^2} \tag{5}$$

and similarly,

$$\tilde{\chi}_H = C_Z \frac{\tilde{\varepsilon}}{\tilde{\kappa}} \widetilde{H''^2} \tag{6}$$

where $\tilde{\kappa}$ and $\tilde{\varepsilon}$ are the Reynolds-averaged turbulent kinetic energy and dissipation rate, and the constant of proportionality ($C_Z$ and $C_H$) represents the appropriate ratio of turbulent and scalar mixing timescales, which is generally assumed to be 2.0. The validity of this model in the LTC conditions is assessed by an *a priori* test using the DNS data. Considering that the DNS configuration is periodic in all directions, the averaging over the entire domain is considered a reasonable representation of the Reynolds-averaging.

Figure 6 shows the temporal variation of the model constants, $C_Z$ and $C_H$, for the three cases simulated. Case A has temperature fluctuations only, and $C_H$ is the only relevant quantity. It is seen that both quantities are nearly constant in the non-reacting regime for all three cases (except for the initial rise which is due to turbulent straining of the mixing field, as an artifact of the initial turbulence seed), although the absolute magnitude of the constant is different for each case, for $Z$ and $H$. The results show that $C_H$ is consistently larger than $C_Z$ by a factor of almost 2-3.

Once the ignition starts, however, $C_Z$ and $C_H$ are no longer constants and exhibit an abrupt rise, more prominently so for Cases A and B for which ignition front formation is more pronounced. This poses a challenge in predictive simulation of the full-cycle RANS simulations of the LTC process, which may lead to inaccurate prediction of ignition delay and subsequent combustion phasing.

Considering that the model constants represent the timescale ratios, a logical step to investigate this issue is to examine the individual time scales for the mixing ($\tau_Z$, $\tau_H$) and turbulence ($\tau_{turb}$) for all the three cases. Figure 7 shows the evolution of the three time scales along with the logarithm of the inverse of integrated heat release rate as a marker of the ignition event. First, the turbulence time scale is seen to increase monotonically, as expected for the decaying turbulence in the present study. The initial drop in the mixing time scales for both $Z$ and $H$ results from the effect of the initial transient associated with turbulent straining of the mixing field, after which the scalar mixing time scales also gradually increase over time. A notable point is that the scalar mixing time scales abruptly drops as the ignition event starts, since the fronts create strong scalar gradients as discussed earlier. From this point onwards until the global volumetric ignition is reached, the mixing timescales are small, and finally increase marginally towards the end of combustion when the spatial gradients within the domain no longer exist.

As discussed in Section 3.1.1, the abnormal behavior is attributed to the mixture fraction gradient formation due to the combined effect of ignition front and differential diffusion. As such, the model constants are also computed from the contrived DNS with the equi-diffusion model, and the results are compared in Figure 8. It is clearly seen that all model constants, for all three cases, remain nearly unchanged throughout the ignition and combustion events. Therefore, the results suggest that the flamelet model must properly account for the differential diffusion effect, especially for the mixture with significantly light or heavy fuels. A proposed approach is to newly define an equi-diffusive mixture fraction variable and derive general reactive scalar equations accounting for differential diffusion [26]. Furthermore, under the unity Lewis number assumption, often adopted in many flamelet models applied to complex engine simulations, the magnitude of the model constants for both $Z$ and $H$ are close to 3.0 instead of 2.0 as commonly used.

A general two-dimensional flamelet approach also involves a Reynolds-averaged cross-scalar dissipation rate term $\tilde{\chi}_{ZH} = 2\alpha |\nabla Z||\nabla H|$. Based on the above findings, a simple heuristic model is proposed as:

$$\tilde{\chi}_{ZH} = C_{ZH} \frac{\tilde{\varepsilon}}{\tilde{\kappa}} \left( Z''^2 H''^2 \right)^{1/2} \tag{7}$$

Figure 9 shows the value of $C_{ZH}$ for Cases B and C, based on the equi-diffusion model. The value is approximately 3.0, and the validity of Equation (7) is justified.

### 3.1.3. Conditional scalar dissipation rates

The scalar dissipation rates conditioned for the $Z$ and $H$ variables, denoted as $\langle \chi_Z | Z \rangle$ and $\langle \chi_H | H \rangle$, respectively, appear in the flamelet equations, and provide the effect of mixing on the reactive scalar evolution. These terms are commonly modeled as [11, 27]:

$$\langle \chi_Z | Z \rangle = \frac{\tilde{\chi}_H f(Z)}{\int_0^1 f(Z) P(Z) dZ} \tag{8}$$

where *P(Z)* is the PDF of *Z* and *f(Z)* represents the functional dependence of scalar dissipation rate on *Z*. In general, *f(Z)* is modeled either as a 1D infinite mixing layer or as a counterflow flame [11]. Here the results for the 1D infinite mixing layer model are presented. The results based on the counterflow flame model are nearly identical and are not reported here. Figure 10 shows the conditional scalar dissipation rate profile at various instants for Case B, determined directly from the DNS data and by the model. During the ignition front propagation (1.5 ms to 2.5 ms), the DNS-based results show considerably higher peak values that are shifted in the *Z* axis over time. The increase in peak values is due to the increase in gradients of *Z* and *H* during ignition front propagation. As discussed in Section 3.1.1, this is again attributed to the differential diffusion effect. Using the equi-diffusion model DNS as shown in Figure 11, the high peaks are substantially suppressed. However, there still exist high peaks in the large *Z* regions, which are not captured by the model. This suggests that a different functional form may be needed to accurately capture the conditional scalar dissipation rate under HCCI conditions. Formulation of such an appropriate profile is the subject of future work.

### 3.2. Assessment of LES Mixing Models

In the previous sections, the Reynolds-averaging is represented by the averaging of the DNS data over the entire domain. Alternatively, the DNS data can be spatially filtered at a filter size less than the domain size for *a priori* tests for LES submodels. An appropriate LES filter size should be small enough to capture the energy containing turbulent eddies but larger than the scales to characterize the scalar dissipation rate [28]. Figure 12 shows a typical snapshot of the scalar dissipation rate field, taken from Case B at 2 ms. The thickness of the scalar dissipation rate profile is approximately 0.06 mm. Hence, a filter size of 30 times the DNS grid size (0.13 mm), which is approximately twice the scalar dissipation rate scale, is adopted in this analysis. The mixing model in LES has commonly adopted the same RANS formula, Equations (5) and (6) with Reynolds-averaging substituted by spatial filtering. The validity of such an approach is assessed here. For convenience of notation, the spatial LES filtered variables are denoted by the same tilde, such as $\tilde{Z}$.

First, Figure 13 shows the correlation between $\tilde{\varepsilon}/\tilde{k}$ and $\tilde{\chi}_H$ computed from the DNS data for Case B. It is evident that the turbulence quantity is not at all correlated with the filtered scalar dissipation rate. Considering that LES filter size is smaller than large eddy scale, this result implies that the subgrid turbulence time scales does not directly affect the mixing characteristics. Instead, much stronger correlations are found by using an alternative model formula such as:

$$\tilde{\chi}_Z = C\widetilde{Z'''^2}, \quad \tilde{\chi}_H = C\widetilde{H'''^2}, \quad \tilde{\chi}_{ZH} = C\left(\widetilde{Z'''^2}\,\widetilde{H'''^2}\right)^{1/2} \qquad (9)$$

with $C = 10^{4.2}$ (1/s). Figure 14 shows the correlation for the three relations for Case B. The results demonstrate that at the LES filter level more direct correlations are found between the filtered variance and the filtered scalar dissipation rate. A similar finding was reported in IC engine LES simulations [Ref 29: Rutland]. The results are also found to be the same for test cases with unity Lewis number assumption, suggesting that differential diffusion effect is not important in this case.

## 4. CONCLUSIONS

Two-dimensional direct simulations representing LTC ignition conditions were utilized in an *a priori* test of RANS and LES mixing models for flamelet formulation. The simulation cases represent different combustion scenarios, in which ignition is dominated by front propagation or by nearly homogeneous explosion.

For the RANS modeling, three components of the existing models were examined. First, the beta-PDF representation of the mixture fraction distribution was found to be valid, although strong differential diffusion effects may induce some unphysical behavior of the magnitude and shift in the peak PDF values. Secondly, the proportionality constant for the Reynolds-averaged scalar dissipation rate was also found to be strongly affected by the differential diffusion effect during the ignition event. These findings were also confirmed by test simulations employing an equi-diffusion model. In the equi-diffusion cases, the model constants remain unaffected by ignition, while the quantitative value is found to be close to 3.0 rather than 2.0, which has been commonly adopted in previous studies. Lastly, the conditional scalar dissipation statistics showed some discrepancies compared to the existing model based on laminar flamelet assumption for the presumed $Z$-dependence, and this issue needs to be addressed in the future work. All together, the differential diffusion effect, when relevant, was found to be a significant factor, and needs to be properly accounted for in the flamelet formulation framework.

An *a priori* test for LES mixing models was also conducted by filtering the DNS data at an appropriate filter size. The results showed that a more direct correlation between the scalar dissipation and the filtered variance is found, while the correlation with filtered turbulence time scales was found to be weak. A simpler model formulations were suggested for $Z$, $H$, and cross-scalar dissipation rates, and a universal constant was found to be sufficient to lead to good correlations.

## 5. ACKNOWLEDGMENTS

computational resources for the 2-D DNS simulations were supported in part by the National Science Foundation through TeraGrid provided by Pittsburgh Supercomputing Center. The authors would like to thank Dr. Gaurav Bansal for providing the simulation data. HGI was affiliated with University of Michigan while the main research activities were undertaken.

[29]

**FIGURES (WITH CAPTIONS)**

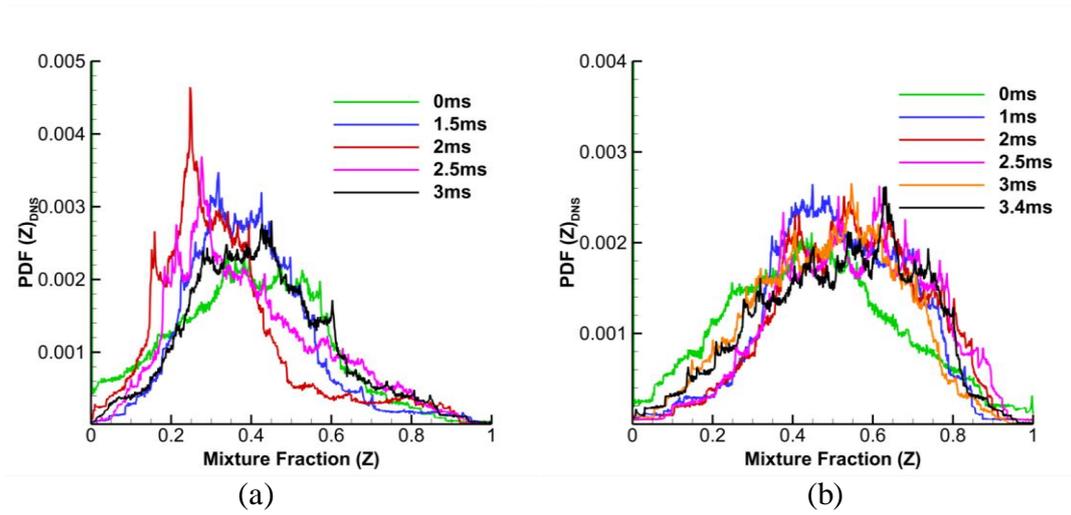

(a)　　　　　　　　　　　　　　(b)

Figure 1: PDF of mixture fraction at different times during the evolution of simulation, calculated from the DNS data, for (a) Case B and (b) Case C.

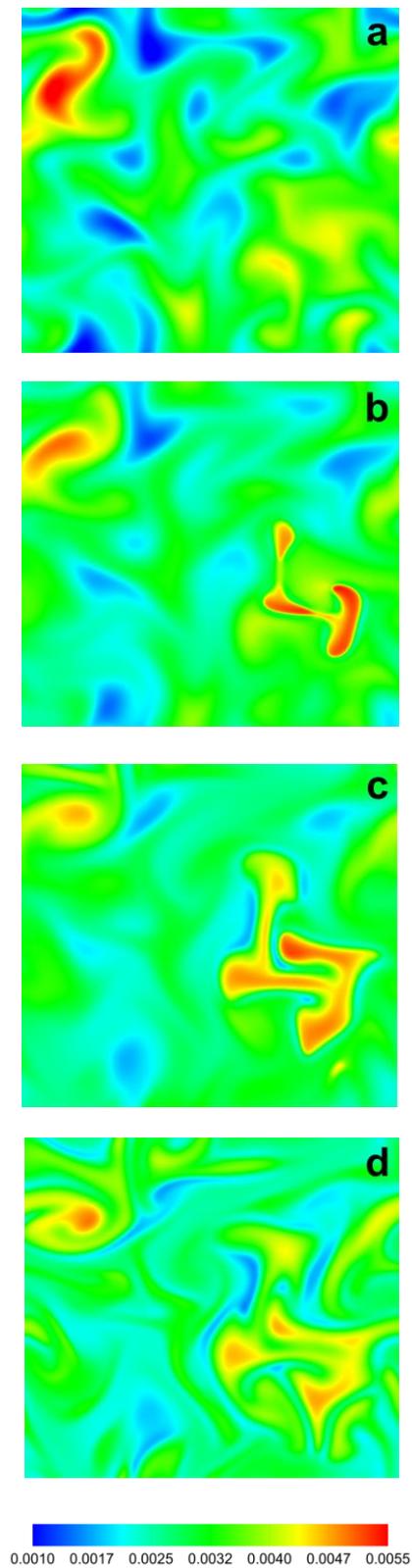

Figure 2: The mixture fraction field evolution for Case B: at (a) $t = 1.0$ ms, (b) $t = 1.5$ ms, (c) $t = 2.0$ ms, and (d) $t = 2.5$ ms.

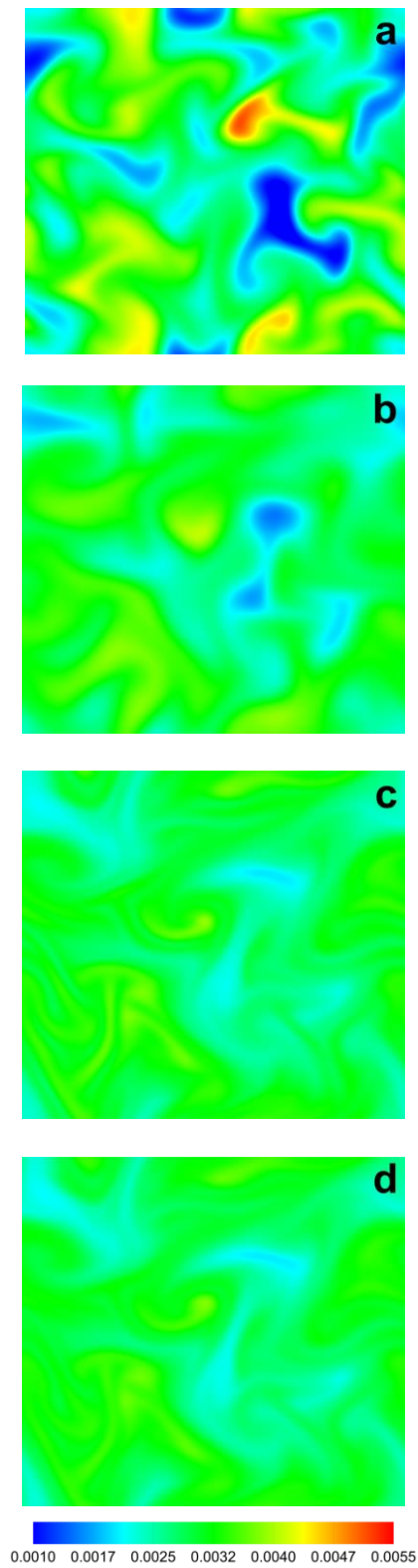

Figure 3: The mixture fraction field evolution for Case C: at (a) $t = 1.0$ ms, (b) $t = 2.0$ ms, (c) $t = 3.0$ ms, and (d) $t = 3.2$ ms.

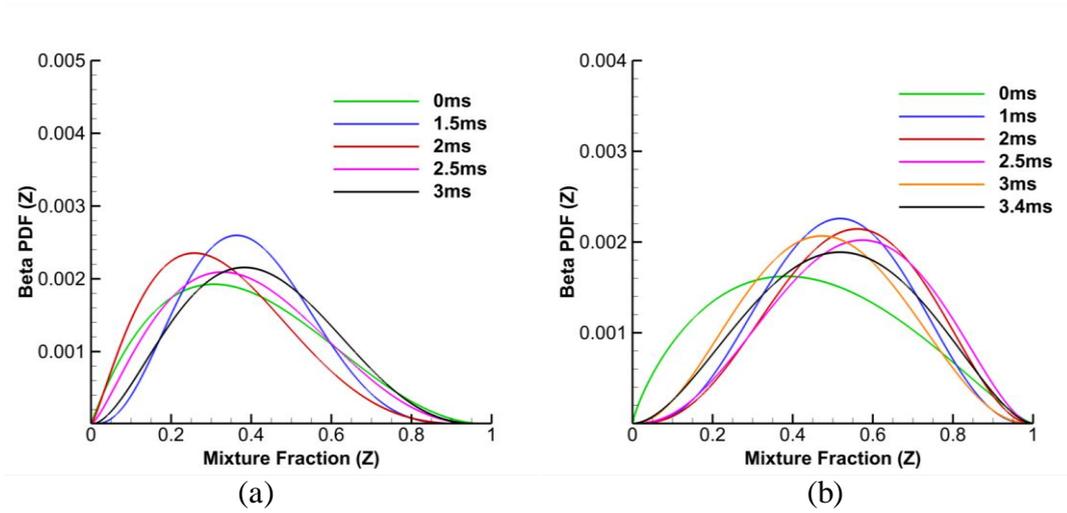

Figure 4: PDF of mixture fraction at different times during the evolution of simulation determined by the beta-PDF model, for (a) Case B and (b) Case C.

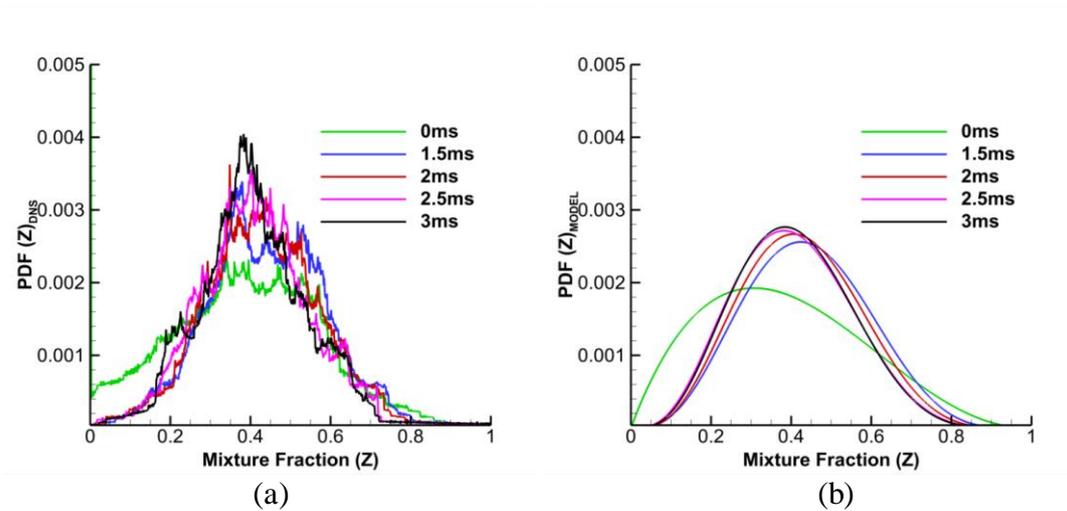

Figure 5: PDF of mixture fraction at different times for Case B corresponding to Figure 1, but with the equi-diffusion model, (a) calculated from DNS data, and (b) calculated from the beta-PDF model.

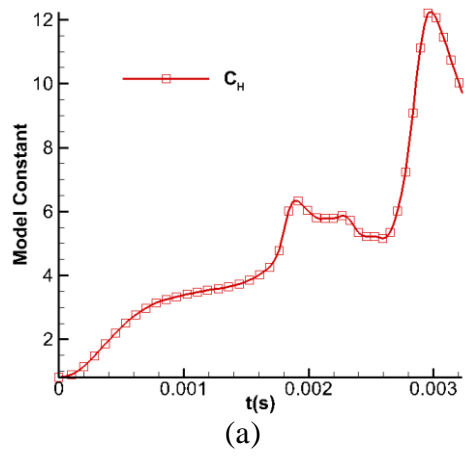

(a)

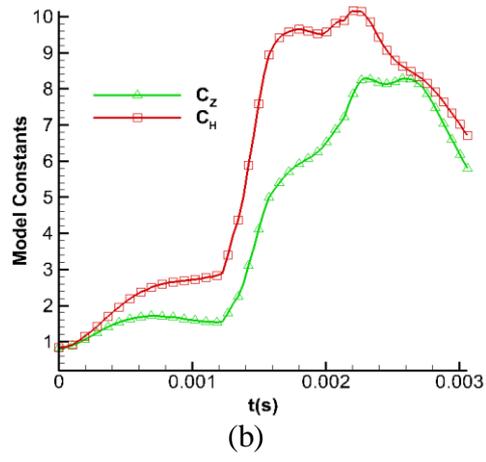

(b)

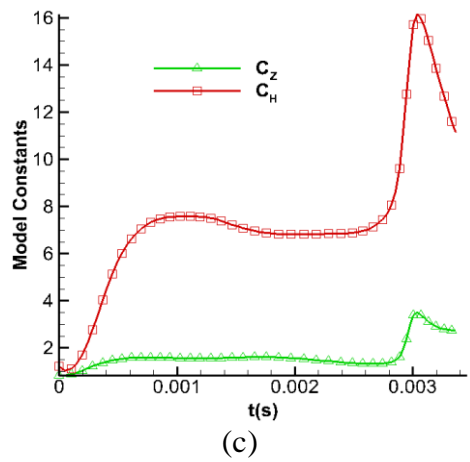

(c)

Figure 6: The evolution of the model constants, $C_Z$ and $C_H$, for (a) Case A, (b) Case B, and (c) Case C.

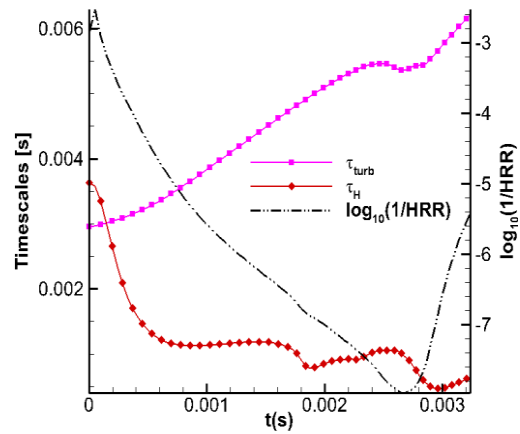

(a)

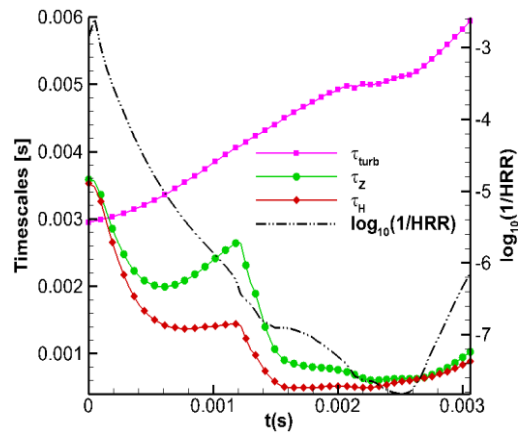

(b)

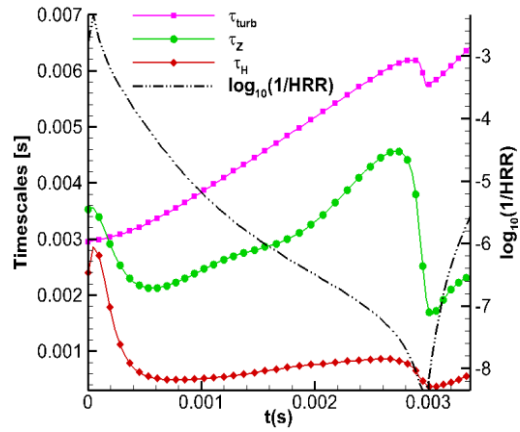

(c)

Figure 7: The evolution of time scales for turbulence ($\tau_{turb}$), $Z$ mixing ($\tau_Z$), and $H$ mixing ($\tau_H$), for (a) Case A, (b) Case B, and (c) Case C. The heat release time scale is also overlaid.

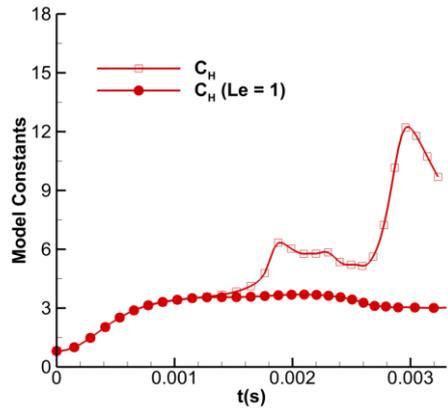

(a)

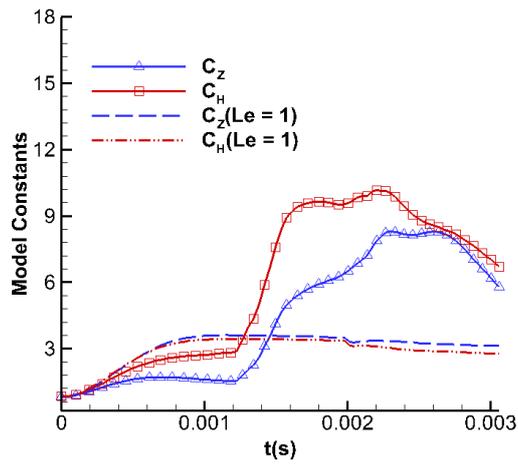

(b)

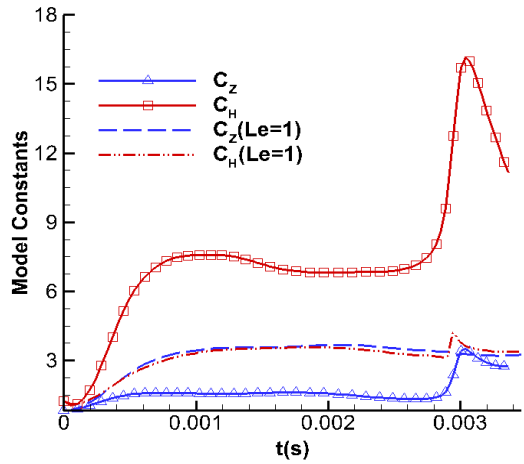

(c)

Figure 8: Comparison of the evolution of the model constants $C_Z$ and $C_H$, shown in Figure 6 (squares) and the results from DNS with the equi-diffusion model (circles), for (a) Case A, (b) Case B, and (c) Case C.

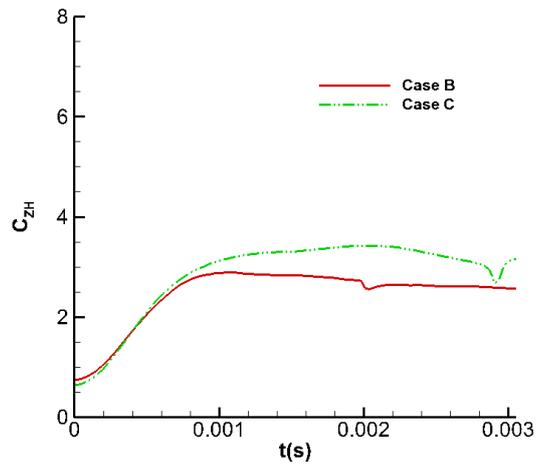

Figure 9: The evolution of the model constant for the cross scalar dissipation rate, $C_{ZH}$, determined with the equi-diffusion model for Cases B and C.

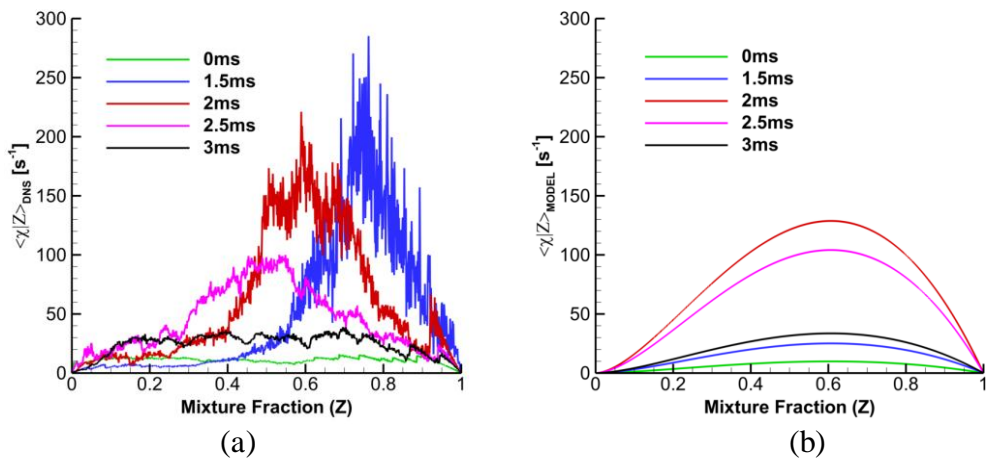

(a)          (b)

Figure 10: The conditional scalar dissipation rate, $\langle \chi | Z \rangle$, at different times for Case B, (a) determined from the DNS data, and (b) determined by the 1D mixing layer model.

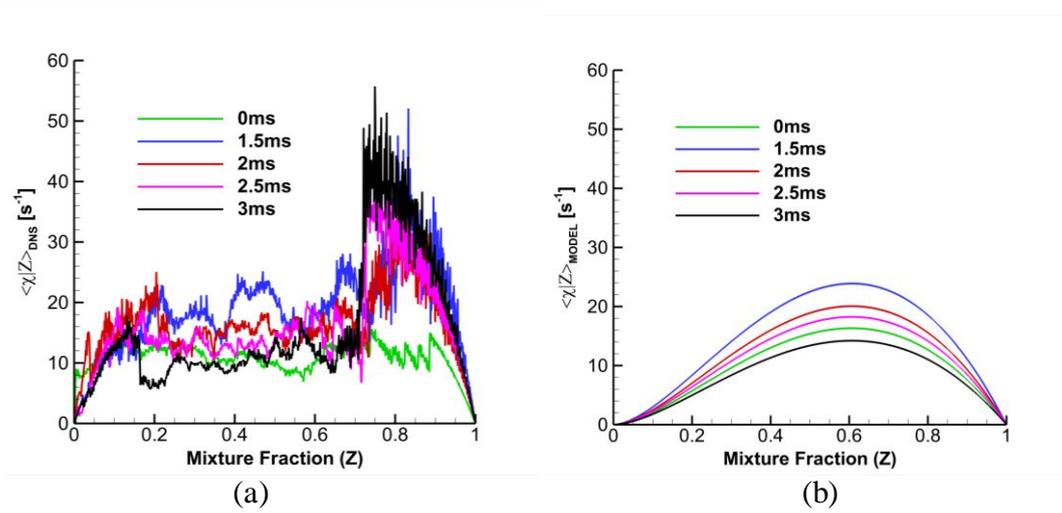

Figure 11: The conditional scalar dissipation rate, $\langle \chi | Z \rangle$, at different times for Case B for DNS with the equi-diffusion model, (a) determined from the DNS data, and (b) determined by the 1D mixing layer model.

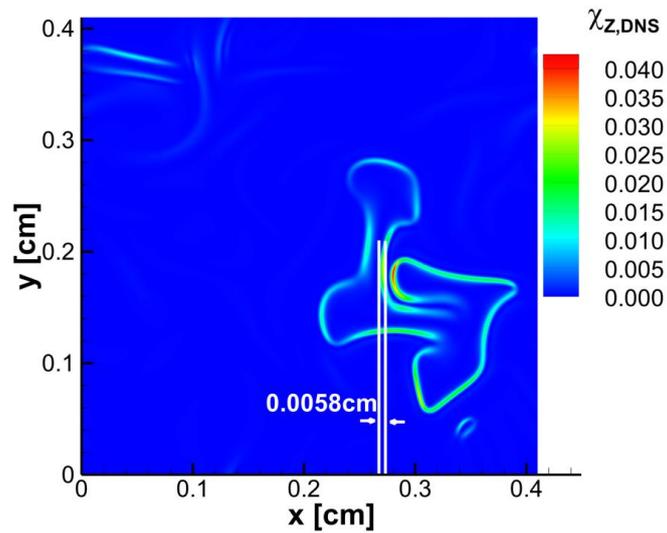

Figure 12: The spatial profiles of the $Z$ scalar dissipation rate for Case B at $t = 2$ ms, showing the characteristic thickness.

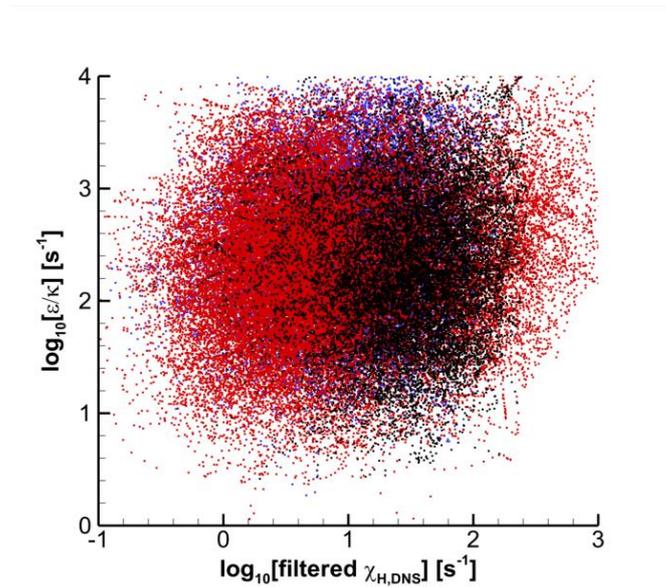

Figure 13: Correlation between the inverse eddy turnover time, $\tilde{\varepsilon}/\tilde{k}$, and the filtered $H$ scalar dissipation rate, $\tilde{\chi}_H$, for Case B (filter size = $30\Delta_{DNS}$).

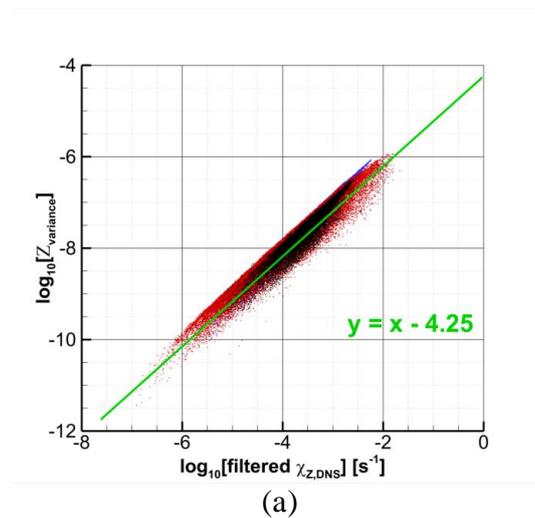

(a)

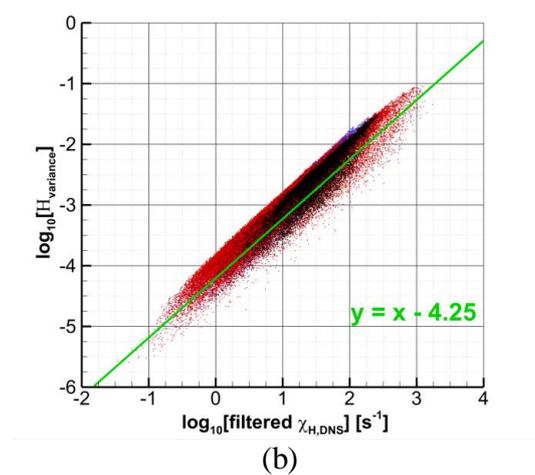

(b)

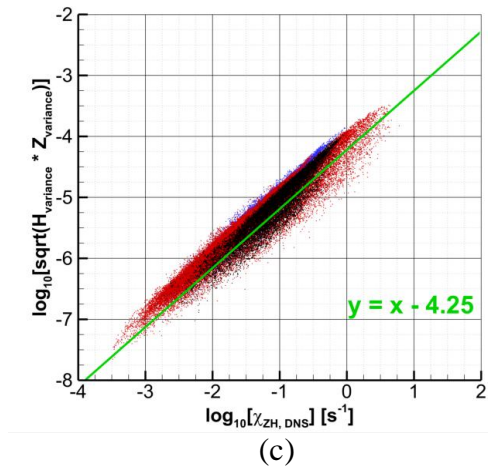

(c)

Figure 14: Correlation between the filtered scalar dissipation rate and the filtered variance for (a) *Z*, (b) *H*, and (c) *Z-H*, determined for Case B (filter size = $30\Delta_{DNS}$).